\documentclass{JHEP}

\newcommand{\Xgt}{\Upsilon}
\newcommand{\Or}{{\cal O}}
\newcommand{\eps}{\epsilon}
\newcommand{\Nu}{{\cal V}}
\newcommand{\be}{\begin{equation}}
\newcommand{\ee}{\end{equation}}
\newcommand{\bea}{\begin{eqnarray}}
\newcommand{\eea}{\end{eqnarray}}
\newcommand{\bean}{\begin{eqnarray*}}
\newcommand{\eean}{\end{eqnarray*}}

\def\beq{\begin{equation}}

\def\bra#1{\left\langle #1\right|}
\def\eeq{\end{equation}}
\def\half{\frac{1}{2}}

\def\ket#1{\left| #1\right\rangle}

\def\crbig{\\\noalign{\vspace {3mm}}}
\relax

\preprint{MIT-CTP-3203\\  {\tt hep-th/0110204}}

\title{Some exact results on the matter star-product in the 
half-string formalism}

\author{Nicolas Moeller
\\
Center for Theoretical Physics,
\\Massachusetts Institute of Technology,\\ Cambridge, MA 02139, USA\\
\email{moeller@mit.edu}
}

\abstract{
We show that the D25 sliver wavefunction, just as the D-instanton sliver, 
factorizes when expressed in terms of half-string coordinates. 
We also calculate analytically the star-product of two zero-momentum eigenstates of 
$\hat{x}$ using the vertex in the oscillator basis, thereby showing that the 
star-product in the matter sector can indeed be seen as multiplication of matrices 
acting on the space of functionals of half strings.
We then use the above results to establish that the matrices 
$\rho_{1,2}$, conjectured by Rastelli, Sen and Zwiebach to be left and right 
projectors on the sliver, are indeed so.
}
\begin{document}

\section{Introduction} \label{s0}

It has early been proposed \cite{WITTENSFT} that the Open String Field Theory 
star-product may be 
seen as an infinite dimensional local matrix algebra. Namely, if one writes 
a string functional $\psi$ as a functional $\psi(x_{\rm mid}, x^L, x^R)$ of the 
string midpoint $x_{\rm mid}$, the left-half modes $x^L$ and the right-half 
modes $x^R$, the star-product of two string functionals sharing the same 
midpoint could then be written
\beq
(\psi_1 \star \psi_2)(x_{\rm mid}, x^L, x^R) = 
\int{[d y] \, \psi_1(x_{\rm mid}, x^L, y) \, \psi_2(x_{\rm mid}, y, x^R)} \,,
\label{introstar}
\eeq
where $y$ is a half-string coordinate.

\paragraph{}
This operator formulation of the star-product had an early important application. 
Indeed, it was implicitly used by Gross and Jevicki in \cite{GJ} and \cite{GJ2} 
to deduce the form of 
the star-product in terms of Neumann coefficients; it actually provided  
expressions of these Neumann coefficients without using the conformal mapping of 
the vertex. One of the results of the present paper is to take the reverse path: 
starting 
from the star-product given in terms of Neumann coefficients, we are able to 
show analytically that, in the zero-momentum matter sector, the star-product 
indeed has the form 
(\ref{introstar}). For this, we carefully calculate the star-product 
$\ket{x_1} \star \ket{x_2}$ of two zero-momentum eigenstates of $\hat{x}$ with 
vanishing zero-mode  
$\left( x_{1,2} \right)_0 = 0$. We find that the result is proportional to 
$\delta(x_1^R - x_2^L) \, \ket{x_3}$, where the left-half of $x_3$ is equal to the 
left-half of $x_1$ and the right-half of $x_3$ is equal to the right-half of $x_2$ 
(namely $x_3^L = x_1^L$ and $x_3^R = x_2^R$). The delta function implies that the result 
is zero except when the right-half of $x_1$ coincides with the left-half of $x_2$. 
This calculation is technical and will, moreover, require the use of a 
regularization scheme. At last, using the fact that the states $\ket{x}$ form a complete 
basis of zero-momentum states, we can formally derive the form of the star-product 
(\ref{introstar}) between zero-momentum string wavefunctions.

\paragraph{}
Early attempts have been made to develop the operator formalism of the star-product 
(\cite{Chan-Tsou} - \cite{Abdurrahman-Bordes}). And more recently, this  
formalism gained new importance in the context of Vacuum String Field Theory, 
proposed by Rastelli, Sen and Zwiebach in (\cite{RSZ} - \cite{RSZboundary}) 
(see \cite{RSZvacuum} for a nice review).
In this proposal, the BRST operator around the stable vacuum is taken to be pure ghost, 
and thus the equation of motion around the 
nonperturbative vacuum factorizes into 
a matter part and a ghost part. In particular in the matter part, the equation of 
motion takes the form of a projection equation:
\beq
\psi_{\rm m} \star \psi_{\rm m} = \psi_{\rm m} \,.
\label{introproj}
\eeq

\paragraph{}
The simplest nontrivial zero-momentum solution of this equation, besides the 
identity, is the renowned sliver, which 
was first constructed in its geometric form by Rastelli and Zwiebach in \cite{RZ} 
as the limit 
$n \rightarrow \infty$ of the wedge states $\ket{n}$. 
Kostelecky and Potting \cite{KP} later found an algebraic solution of (\ref{introproj}), 
which was then conjectured by Rastelli, Sen and Zwiebach \cite{RSZclassical} to be 
the same as the geometric 
sliver. A nice and almost complete proof of this conjecture was given recently 
by Furuuchi and Okuyamain in \cite{Comma}.
It was also conjectured in \cite{RSZclassical} that the sliver solution corresponds 
to a space-filling D-25 brane, bringing us back from the stable vacuum to 
the perturbative vacuum.

\paragraph{}
An interesting property of the sliver was proposed in \cite{RSZhalf}: The authors 
found numerical evidence that, when expressed in terms of half-string modes, 
the sliver wavefunction seems to factorize into a left part and a right part. 
Their motivation was that, when the sliver is interpreted as the wedge state 
$\ket{\infty}$, the surface describing the sliver state is seen, in a suitable 
coordinate system, 
to be cut in two disjoint pieces: one corresponding to the left degrees of freedom 
and one corresponding to the right degrees of freedom (see \cite{RSZboundary} for a 
detailed treatment of the geometric picture).
Gross and Taylor \cite{GT} independently found a similar property regarding the 
D-instanton sliver (a sliver localized in all 26 dimensions), and they were able 
to prove it. Their proof is based on the use of the overlap equations satisfied 
by the Neumann coefficients. These equations 
were formally shown to hold in (\cite{GJ}, \cite{GJ2}) for the Neumann coefficients 
found from the conformal 
mapping of the vertex, and they were interpreted as the condition that the vertex 
glues the right-half of the $r^{\rm th}$ string to the left-half of the 
$(r+1)^{\rm th}$ string. These equations thus implicitly contain information about 
the half-string formalism, which is necessary for the proof.

\paragraph{}
The proof that the D-instanton sliver wavefunction factorizes however doesn't 
immediately generalize to the sliver. It is 
one of the aims of this paper to present a proof of the factorization of the sliver.
It is mainly based on the proof of Gross and Taylor, but requires some modifications 
due to the nontrivial transformations of the Neumann coefficients when going from 
the zero-momentum basis to the momentum-dependent basis.
This factorization property is important in the projection operator techniques 
used in \cite{RSZhalf} and \cite{GT}. Indeed, if we write the sliver wavefunction as 
$\psi_{\Xi}(x^L, x^R) = f(x^L) \, f^*(x^R)$, 
we see that it is a rank one projector on the space of half-string functionals.

\paragraph{}
Finally, we use our previous results to show that the operators $\rho_{1,2}$ 
introduced by Rastelli, Sen and Zwiebach, satisfy some equations which allow us to 
interpret them as left and right projectors on the sliver, as conjectured 
in \cite{RSZhalf}.

\paragraph{}
The outline of the paper is as follows. In section \ref{s1} we introduce our notation 
and prove that the sliver factorizes. In the course of the proof, we will derive several 
useful equations that will be used in section \ref{s3}. In section \ref{s2}, we set up 
the calculation of $\ket{x_1} \star \ket{x_2}$ and we motivate it by giving an 
application of 
its result: we show that the operators $\rho_1$ and $\rho_2$ of \cite{RSZhalf} are indeed 
left and right projectors on the sliver. In section \ref{s3}, we complete the proof 
started in section \ref{s2}. Then in section \ref{s4} we collect 
some useful results established in this paper. And finally, in section \ref{s5}, 
we discuss our results and propose further directions of investigation.

\section{Factorization of the sliver} \label{s1}

In \cite{GT}, the authors proved that the D-instanton sliver factorizes when 
its wavefunction is expressed in terms of half-string modes. Using a similar 
proof, we show here that the sliver (that we sometimes call the 
D-25-brane sliver to avoid confusion) constructed in \cite{RSZclassical}, 
factorizes as well.

\subsection{Notations}
Our notations are taken from different sources (in particular \cite{RSZclassical}, 
\cite{GT}, \cite{Comma}). And as the same symbols are sometimes used for different 
quantities in different papers, we start this section by giving an exhaustive list 
of our own notation.

\subsubsection*{Neumann coefficients}
The star-product of two states $\ket{A}$ and $\ket{B}$ is calculated as 
\beq
\ket{A} \star \ket{B}_3 =  ~_1\bra{A}~_2\bra{B} V_3 \rangle_{123} \,,
\eeq
where $\ket{V_3}$ is the three-string vertex. We can write it explicitly 
in terms of Neumann coefficients, which will depend on the basis we choose. 
In the following, we will use two different bases: The {\em zero-momentum basis}, in 
which
\beq
\ket{V_3}_{123} = \exp \left( -\half \sum_{r, s \atop m, n \geq 1} a^{(r)\dagger}_m 
\cdot V^{r s}_{m n} a^{(s)\dagger}_n \right) \ket{0}_{123} \,,
\eeq
where the state $\ket{0}$ (the zero-momentum vacuum whose zero-mode part has 
been dropped) is annihilated by the $a_n$'s for $n \geq 1$, and satisfies 
$\langle 0 | 0 \rangle = 1$.
Note that a summation over Lorentz indices is implicit in the $\cdot$ product. 
The $V^{rs}$ are the Neumann coefficients and the superscripts $r$ and $s$, labeling 
the Hilbert space, go from one to three. Note also that the modes $m$, $n$ run from 
one to infinity, and the states $\ket{A}$ and $\ket{B}$ star-multiplied using this 
vertex, must have zero momentum.

\paragraph{}
We will also use our {\em momentum-dependent basis}, in which
\beq
\ket{V_3}_{123} = \exp \left( -\half \sum_{r, s \atop m, n \geq 0} a^{(r)\dagger}_m 
\cdot \tilde{\cal V}^{r s}_{m n} a^{(s)\dagger}_n \right) \ket{\Omega}_{123} \,,
\eeq
where $a_0 = \half (p_0 - 2 i \, x_0)$ obeys the commutation relation 
$[a_0, a_0^\dagger] = 1$, and the vacuum $\ket{\Omega}$, satisfying 
$\langle \Omega | \Omega \rangle = 1$, is annihilated by all the $a_n$'s 
including $a_0$, it is therefore 
different from $\ket{0}$. Note also that here the modes 
$m$ and $n$ run from zero to infinity, and the states $\ket{A}$ and $\ket{B}$ 
star-multiplied in this basis are general.  
This particular momentum-dependent basis was used in \cite{GT} to construct the 
D-instanton sliver, and the zero-momentum basis was earlier used to describe the 
sliver (\cite{KP}, \cite{RSZclassical}). 

\paragraph{}
It is time for some clarification:
Matrices describing the D-instanton sliver have indices running from $0$ to 
$\infty$, but the indices of the matrices used for the D-25-sliver run only 
from $1$ to $\infty$. For clarity we will thus denote matrices whose indices 
start from zero with a $\tilde{ }$ such as $\tilde{M}$ , as matrices 
without a $\tilde{ }$ will have indices starting from one. If a matrix 
$\tilde{M}$ is defined, we will always assume that the corresponding 
matrix $M$ is obtained from $\tilde{M}$ by deleting its first column and its 
first row.
Moreover, the matrices transforming non-trivially from one basis to the 
other (like $V^{r s}$) will be written in calligraphic form in the momentum-dependent 
basis, and in roman form in the zero-momentum basis.

\paragraph{}
The matrices $V^{r s}\ (\tilde{\cal V}^{r s})$ satisfy the following 
transposition and cyclicity properties:
\beq
\left(V^{r s} \right)^T = V^{s r} \,, \qquad 
V^{r s} = V^{(r+1) (s+1)} \,.
\eeq
They can be written in terms of a single 
matrix $U\ (\tilde{\cal U})$:
\beq
V^{r s} = {1 \over 3} \, \left( C + \omega^{s-r} \, U + \omega^{r-s} \, \bar{U} \right) \,,
\qquad
\tilde{\cal V}^{r s} = {1 \over 3} \, \left( \tilde{C} + \omega^{s-r} \, \tilde{\cal U} + 
\omega^{r-s} \, \bar{\tilde{\cal U}} \right) \,,
\label{VfromU}
\eeq
where $\tilde{C}_{m n} = (-1)^m \, \delta_{m n}$ and $\omega = e^{2i \pi/3}$. 
The matrices $U$ and $\tilde{\cal U}$ obey the following relations:
\beq
\bar{U} \equiv U^* = C U C \,, \quad U^2 = \bar{U}^2 = 1 \,, \quad U^\dagger = U \,, 
\quad \bar{U}^\dagger = \bar{U} \,,
\label{Uprop}
\eeq
as well as similar equations for $\tilde{\cal U}$. The relation between $U$ and 
$\tilde{\cal U}$ is given by (\cite{KP}, \cite{RSZclassical})
\beq
U_{mn} = \tilde{\cal U}_{mn} + {\tilde{\cal U}_{m0} \, \tilde{\cal U}_{0n} 
\over 1 - \tilde{\cal U}_{00}} \,.
\label{Urel}
\eeq
We now define the matrices $X$, $Y$ and $Z$:
\beq
X = C V^{11} \,, \quad Y = C V^{12} \,, \quad Z = C V^{21} \,.
\eeq
They are symmetric and satisfy the relations:
\bea
&& X + Y + Z = 1 \,,
\nonumber \\
&& X^2 + Y^2 + Z^2 = 1 \,,
\nonumber \\
&& Y Z = X^2 - X \,,
\nonumber \\
&& C X C = X \,, \quad C Y C = Z \,, \quad C Z C = Y \,.
\eea

\subsubsection*{Half-string formalism}
Let us now review the half-string formalism as it was presented in \cite{RSZhalf}. Let 
us first write the mode expansion of the open string coordinate:
\beq
X^\mu(\sigma) = x_0^\mu + \sqrt{2} \sum_{n=1}^\infty x_n^\mu \, \cos (n \sigma) \,,
\quad 0 \leq \sigma \leq \pi \,.
\eeq
We then define the left and right half of the string as:
\bea
X^{L \mu}(\sigma) &=& X^{\mu}(\sigma/2) - X^\mu(\pi/2) \,, \, \ \ \qquad 
0 \leq \sigma \leq \pi \,,
\nonumber
\\
X^{R \mu}(\sigma) &=& X^{\mu}(\pi - \sigma/2) - X^\mu(\pi/2) \,, \quad  
0 \leq \sigma \leq \pi \,.
\eea
They have the following mode expansion:
\beq
X^{L (R) \mu}(\sigma) = \sqrt{2} \, \sum_{n=1}^\infty x_n^{L (R) \mu} \, 
\cos\left( \left( n - \half \right) \, \sigma \right) \,.
\eeq
We can express the relation between the full modes and the half modes in terms of 
matrices $A^\pm$ (from now on, we will not write the spacetime indices, which will 
be implicit):
\beq
x = A^+ x^L + A^- x^R \,,
\label{fullfromhalf}
\eeq
where 
\beq
A^\pm_{n m} = \pm \half \delta_{n, 2m-1} + {1 \over 2 \pi} \, (1 + (-1)^n) \, 
(-1)^{m + \half n - 1} \, \left( {1 \over 2m + n -1} + {1 \over 2m - n -1} 
\right) \,.
\eeq
We can invert the relation (\ref{fullfromhalf}), and we get:
\beq
x^L = \tilde{A}^+ x \,, \qquad x^R = \tilde{A}^- x \,,
\eeq
where\footnote{Note that the relation between $A$ and $\tilde{A}$ has nothing 
to do with the notation introduced earlier; we want here to keep the notations 
of \cite{RSZhalf}. The indices of both $A$ and $\tilde{A}$ run from one to infinity. 
We hope that the reader will not get confused with this slight abuse of notation.}
\beq
\tilde{A}^\pm_{m n} = 2 \, A^\pm_{n m} - {1 \over \pi} \, (1 + (-1)^n) \, 
(-1)^{m + \half n - 1} \, {2 \over 2m -1} \,.
\eeq
Let us note the following useful relations:
\beq
A^+ \tilde{A}^+ + A^- \tilde{A}^- = 1 \,, \quad \tilde{A}^\pm A^\mp = 0 \,, 
\quad \tilde{A}^\pm A^\pm = 1 \,,
\label{Arels}
\eeq
as well as
\beq
A^+ = C A^- \,, \qquad \tilde{A}^+ = \tilde{A}^- C \,.
\eeq
It will be useful to use the matrix $\tilde{\Xgt}$, defined by\footnote{This matrix 
is called $X$ in \cite{GT}, but here we keep the symbol $X$ for  
the matrix $X = CV^{11}$.}
\bea
\tilde{\Xgt}_{2k+1, 2n} = \tilde{\Xgt}_{2n, 2k+1} &=& 
{4 \, (-1)^{k+n} (2 k +1) \over \pi \, ((2 k + 1)^2 - 4 \, n^2)} \qquad (n \neq 0) \,,
\nonumber
\\
\tilde{\Xgt}_{2k+1, 0} = \tilde{\Xgt}_{0, 2k+1} &=& {2 \sqrt{2} (-1)^k \over 
\pi \, (2 k + 1)} \,,
\label{upsdef}
\eea
all other elements being zero.
We can write a relation between $A^\pm$ and $\Xgt$: 
\beq
A^{\pm}_{n m} = \pm \half \, \delta_{n,2m-1} + \half \, \Xgt_{n, 2m-1} \,, \quad n,m 
\geq 1 \,.
\label{AX}
\eeq
Finally, we note the relation between the oscillators $a_n$ and the modes $x_n$:
\beq
\hat{x} = {i \over 2} \, E \cdot (a - a^\dagger) \,,
\eeq
where\footnote{Here we adopt the definition of 
\cite{RSZclassical}, which differs from the one in \cite{GT} by a factor 
of $\sqrt{2}$.}
\beq
\tilde{E}_{nm} = \delta_{nm} \, \sqrt{{2 \over n}} + \delta_{n0} \, \delta_{m0} \,.
\eeq

\subsubsection*{Parity notation}
We will adopt the idea, used in \cite{GT}, to split matrices 
according to the parity of the indices. A matrix $M$ will be rewritten as 
$M = 
\left( \begin{array}{cc} 
M_{oo} & M_{oe}  \\ 
M_{eo} & M_{ee} 
\end{array} \right),$
where the subscripts $o$ and $e$ mean {\em odd} and {\em even} respectively, to denote 
odd and even indices.
With this notation, that we shall call {\em parity notation}, 
the matrix $\tilde{\Xgt}$ can be rewritten 
$\tilde{\Xgt} = \left( \begin{array}{cc} 
0 & \tilde{\Xgt}_{oe}  \\ 
\tilde{\Xgt}_{eo} & 0 
\end{array} \right).$
From the definition (\ref{upsdef}), one can check that 
$\tilde{\Xgt}_{oe} \, \tilde{\Xgt}_{eo} = \tilde{\Xgt}_{eo} \, \tilde{\Xgt}_{oe} = 1$, 
which implies that $\tilde{\Xgt}^2 = 1$; 
when one truncates to matrices with nonzero indices only, $\Xgt_{eo} \, \Xgt_{oe} = 1$ still 
holds, but now $\Xgt_{oe} \, \Xgt _{eo} \neq 1$. In other words, $\Xgt_{eo}$ is a 
left inverse of $\Xgt_{oe}$ but not a right inverse. This will have some 
implications in the way we will establish the factorization.

\paragraph{}
It will be useful in the following to know the form of $X$, $Y$ and $Z$ 
in the parity notation. From (\ref{VfromU}), we get:
\beq
X = {1 \over 3} \, \left( \begin{array}{cc} 1 - 2 \, U_{oo} \;&\; 0 \crbig 
0 \;&\; 1 + 2 \, U_{ee} \end{array} \right) \,,
\ 
Y = {1 \over 3} \, \left( \begin{array}{cc} 1 + U_{oo} \;&\; -\sqrt{3} i \, U_{oe} 
\crbig \sqrt{3} i \, U_{eo} \;&\; 1 - U_{ee} \end{array} \right) \,,
\  
Z = {1 \over 3} \, \left( \begin{array}{cc} 1 + U_{oo} \;&\; \sqrt{3} i \, U_{oe} 
\crbig -\sqrt{3} i \, U_{eo} \;&\; 1 - U_{ee} \end{array} \right) \,.
\eeq

\subsubsection*{The sliver}
The sliver state $\ket{\Xi}$ is given by
\beq
\ket{\Xi} = {\cal N}^{26} \, \exp \left( -\half a^\dagger \cdot S a^\dagger \right) 
\ket{0} \,,
\eeq
where ${\cal N}$ is a normalization factor. And $S = C T$, with
\beq
T = (2 X)^{-1} \, \left( 1 + X - \sqrt{(1 + 3 X) (1 - X)} \right) \,.
\eeq
From the property $S = C S C$, we have that 
$S_{oe} = S_{eo} = 0$.

\vspace{1cm}
\subsection{Proof of the factorization}
In \cite{RSZhalf}, it was shown that the sliver wavefunction, when expressed 
in terms of half-string modes $x^L$ and $x^R$, is
\beq
\psi_{\Xi}(x^L, x^R) = \bra{x} \Xi \rangle = \tilde{\cal{N}}^{26} 
\exp\left(-\half \, x^L \cdot K 
x^L - \half \, x^R \cdot K x^R - x^L \cdot L x^R \right),
\label{psixi}
\eeq
where $\tilde{\cal{N}}^{26}$ is a normalization factor, 
$K = {A^+}^T \, V \, {A^+} = {A^-}^T \, V \, {A^-}$,
$L = {A^+}^T \, V \, {A^-}$, and $V = 2 \, E^{-1} \, {1-S \over 1+S} \, E^{-1}$.
By left-right factorization of the sliver we mean that the cross terms in 
the exponential must vanish. We thus want to show that 
\beq
L \equiv {A^+}^T \, V \, {A^-} = 0
\label{toprove}
\eeq

\paragraph{}
To prove (\ref{toprove}), we first note that (\ref{AX}) implies  the 
following equivalent form of (\ref{toprove}):
\beq
V_{oo} = \Xgt_{oe} \, V_{ee} \, \Xgt_{eo} 
\label{Voo} \,,
\eeq
where we have used that $V_{oe} = V_{eo} = 0$. Now we want to express $V$ in 
terms of $U$. This was done in \cite{GT}, and their result still holds with 
our $U$. Namely, we have:
$$
V \equiv 2 \, E^{-1} \, {1-S \over 1+S} \, E^{-1} = 2 \, E^{-1} \,
\left( \begin{array}{cc} 
\sqrt{3} \, {\sqrt{1-U_{oo}} \over \sqrt{1+U_{oo}}} & 0  \\ 
0 & {1 \over\sqrt{3}} \, {\sqrt{1-U_{ee}} \over \sqrt{1+U_{ee}}}
\end{array} \right) \, E^{-1},
$$
which together with (\ref{Voo}) gives the factorization condition:
\beq
3 \, \sqrt{1-U_{oo} \over 1+U_{oo}} = E \, \Xgt_{oe} \, E^{-1} \, 
\sqrt{1-U_{ee} \over 1+U_{ee}} \, E^{-1} \, \Xgt_{eo} \, E
\label{factorization} \,,
\eeq
which bears some similarities with the factorization condition of the D-instanton 
sliver \cite{GT}.
Our goal now is to prove (\ref{factorization}).

\paragraph{}
For this, we first show that (\ref{factorization}) will hold if the following two 
identities hold:
\bea
\sqrt{3} \, i \, U_{eo} \, (1+U_{oo})^{-1} = E^{-1} \, \Xgt_{eo} \, E \,,
\label{one}
\crbig
-\sqrt{3} \, i \, (1 - U_{oo}) \, (U_{eo})^{-1} = E \, \Xgt_{oe} \, E^{-1} \,.
\label{two}
\eea
Indeed, using the fact that $U^2 = 1$ 
(\ref{Uprop}), we find: 
\beq
\left( \begin{array}{cc} 
U_{oo} U_{oo} + U_{oe} U_{eo} \;&\; U_{oo} U_{oe} + U_{oe} U_{ee} \crbig 
U_{eo} U_{oo} + U_{ee} U_{eo} \;&\; U_{eo} U_{oe} + U_{ee} U_{ee} 
\end{array} \right) 
= 
\left( \begin{array}{cc} 
1 \;&\; 0 \crbig 
0 \;&\; 1 
\end{array} \right) \,.
\eeq
From this, one has
\begin{eqnarray}
& & U_{eo} + U_{ee} U_{eo} + U_{eo} U_{oo} + U_{ee} U_{eo} U_{oo} = 
U_{eo} - U_{ee} U_{eo} - U_{eo} U_{oo} + U_{ee} U_{eo} U_{oo} 
\nonumber \\
& \Rightarrow & (1 + U_{ee}) U_{eo} (1 + U_{oo}) = 
(1 - U_{ee}) U_{eo} (1 - U_{oo}) 
\nonumber \\
& \Rightarrow & {1 + U_{oo} \over 1 - U_{oo}} = 
(U_{eo})^{-1} {1 - U_{ee} \over 1 + U_{ee}} \, U_{eo}
\nonumber \\
& \Rightarrow & \sqrt{1 + U_{oo} \over 1 - U_{oo}} = 
(U_{eo})^{-1} \sqrt{1 - U_{ee} \over 
1 + U_{ee}} \, U_{eo}
\nonumber \\
& \Rightarrow & \sqrt{1 - U_{oo} \over 1 + U_{oo}} = 
(1 - U_{oo}) (U_{eo})^{-1} 
\sqrt{1 - U_{ee} \over 1 + U_{ee}} \, U_{eo} \, {1 \over 1 + U_{oo}}\,,
\label{almostthere}
\end{eqnarray}
where in the fourth line, we assumed that the square root is defined by its 
power expansion. Now substituting (\ref{one}) and (\ref{two}) into 
(\ref{almostthere}) we get exactly (\ref{factorization}).

\paragraph{}
We will now prove the identities (\ref{one}) and (\ref{two}). 
Let us rewrite four equations that were established in \cite{GT} from the overlap 
equations:
\begin{eqnarray}
( 1 + \tilde{\cal U}_{oo}) - \sqrt{3} \, i \, \tilde{E}^{-1} \, 
\tilde{\Xgt}_{oe} \, \tilde{E} \, \tilde{\cal U}_{eo} & = & 0 \ , 
\label{a} \\
- \sqrt{3} \, i \, \tilde{E}^{-1} \, \tilde{\Xgt}_{oe} \, \tilde{E} \, 
( 1 + \tilde{\cal U}_{ee}) + \tilde{\cal U}_{oe} & = & 0 \ , 
\label{b} \\
3 \, ( 1 - \tilde{\cal U}_{oo}) - \sqrt{3} \, i \, \tilde{E} \, 
\tilde{\Xgt}_{oe} \, \tilde{E}^{-1} \, \tilde{\cal U}_{eo} & = & 0 \ , 
\label{c} \\ 
\sqrt{3} \, i \, \tilde{E} \, \tilde{\Xgt}_{oe} \, \tilde{E}^{-1} \, 
( 1 - \tilde{\cal U}_{ee}) - 3 \, \tilde{\cal U}_{oe} & = & 0 \ .
\label{d}
\end{eqnarray}
Multiplying (\ref{a}) by $\tilde{\Xgt}_{eo} \, \tilde{E}$ on the left, one gets
\beq
\tilde{\Xgt}_{eo} \, \tilde{E} \, (1 + \tilde{\cal U}_{oo}) - 
\sqrt{3} \, i \, \tilde{E} \, \tilde{\cal U}_{eo} = 0 \,.
\label{aplus}
\eeq
Our claim is that this equation still holds after replacing ${\cal U}$ by $U$ and 
truncating to nonzero indices:
\beq
\Xgt_{eo} \, E \, (1 + U_{oo}) - \sqrt{3} \, i \, E \, U_{eo} = 0 \,.
\label{preone}
\eeq
To prove this, let us evaluate the $(2n,2m-1)$ component of the lhs of (\ref{preone}), 
for $n, m \geq 1$:
\begin{eqnarray*}
& & \Xgt_{2n,2k-1} \, {1 \over \sqrt{2k-1}} \, (\delta_{2k-1,2m-1} + 
U_{2k-1,2m-1}) - \sqrt{3} \, i \, {1 \over \sqrt{2n}} \, U_{2n,2m-1} 
\\ 
& = & \Xgt_{2n,2k-1} \, {1 \over \sqrt{2k-1}} \, (\delta_{2k-1,2m-1} + 
\tilde{\cal U}_{2k-1,2m-1}) - \sqrt{3} \, i \, 
{1 \over \sqrt{2n}} \, \tilde{\cal U}_{2n,2m-1}+ 
\\
& & + \left( \Xgt_{2n,2k-1} \, {1 \over \sqrt{2k-1}} \, \tilde{\cal U}_{2k-1,0} - 
\sqrt{3} \, i \, {1 \over \sqrt{2n}} \, \tilde{\cal U}_{2n,0} \right) \, 
{\tilde{\cal U}_{0,2m-1} \over 1-\tilde{\cal U}_{00}}
\\ 
& = & \left[ \tilde{\Xgt}_{eo} \, \tilde{E} \, \tilde{\cal U}_{oe} - 
\sqrt{3} \, i \, \tilde{E} \, \tilde{\cal U}_{ee} \right]_{2n,0} \, 
{\tilde{\cal U}_{0,2m-1} \over 1-\tilde{\cal U}_{00}}
\\
& = & \left[ \tilde{\Xgt}_{eo} \, \tilde{E} \, \tilde{\cal U}_{oe} - 
\sqrt{3} \, i \, \tilde{E} \, (1 + \tilde{\cal U}_{ee}) \right]_{2n,0} \, 
{\tilde{\cal U}_{0,2m-1} \over 1-\tilde{\cal U}_{00}}
\\
& = & 0,
\end{eqnarray*}
where in the first step we used (\ref{Urel}), in the 
second step we used (\ref{aplus}), and in the third step we introduced 
an identity matrix which gives no contribution because $\tilde{E}$ is diagonal 
and $n$ is $\geq 1$. And in the last line we used (\ref{b}). 
It is straightforward to show that (\ref{preone}) implies (\ref{one}).

\paragraph{}
Similarly, starting form (\ref{c}) and using (\ref{d}), one can prove (\ref{two}) 
if we assume that the matrix $U_{eo}$ is invertible. This concludes our proof that the 
sliver wavefunction factorizes.

\section{Star-product in the configuration basis} \label{s2}

Here we want to calculate explicitly the star-product $\ket{x_1} \star \ket{x_2}$, where 
$\ket{x_1}$ and $\ket{x_2}$ are zero-momentum eigenstates of $\hat{x}$.
Let us already write down what we expect to find: 
\beq
\ket{x_1} \star \ket{x_2} \sim \delta( x_2^L - x_1^R) \, \ket{x_3}
\label{expect1}
\eeq
where the left-half of $x_3$ is the left-half of $x_1$ and the right-half of $x_3$ is the 
right-half of $x_2$, namely
\beq
x_3 = A^+ x_1^L + A^- x_2^R \, .
\label{expect2} \eeq
The delta function is there because we expect the star-product to be nonzero only 
when the right-half of $x_1$ coincides with the left-half of $x_2$. We will see that 
this delta function indeed arises after regularization of singular terms. 

\paragraph{}
Before proving (\ref{expect1}), we want to motivate its usefulness in the following 
subsection.

\subsection{An application: Left and right projectors on the sliver}

In \cite{RSZhalf}, the authors defined a set of projectors\footnote{The notation 
of \cite{RSZhalf} is a little different of that in 
\cite{RSZclassical} that we are using here: their $Y$ is our $Z$ and their $Z$ 
is our $Y$. Moreover their expressions for the projectors can be simplified by 
use of the relation: $- X Y + Y^2 = Z$ (and similarly: $- X Z + Z^2 = Y$).}$\rho_1$ 
and $\rho_2$:
\beq
\rho_1 = {Z + T Y \over (1+T) (1-X)} \,, \qquad
\rho_2 = {Y + T Z \over (1+T) (1-X)} \,,
\eeq
satisfying
\beq
\rho_1^2 = \rho_1 \,, \quad \rho_2^2 = \rho_2 \,, \quad \rho_1 \rho_2 = 
\rho_2 \rho_1 = 0 \,.
\eeq
They form a complete set of symmetric projectors:
\beq
\rho_1^T = \rho_1 \,, \quad \rho_2^T = \rho_2 \,,  
\quad \rho_1 + \rho_2 = 1 \,.
\eeq
We also note the following relations:
\beq
C \rho_1 C = \rho_2 \,, \qquad \rho_1 - \rho_2 = {Z - Y \over \sqrt{(1-X) 
(1+3 X)}} \,.
\eeq

\paragraph{}
The authors of \cite{RSZhalf} conjectured that $\rho_1$ satisfies 
\beq
\rho_1 \, (1+S) \, E \, \tilde{A}^{+T} = 0 \,,
\label{proj} \eeq
which also implies that $\rho_2 \, (1+S) \, E \, \tilde{A}^{-T} = 0$.
Equation (\ref{proj}) means that $\rho_1$ is a left projector on the sliver in 
the following sense\footnote{I thank B.~Zwiebach for this definition}:
Given a vector $\beta$, the state $(\beta \cdot a^\dagger) \ket{\Xi}$ can be 
written in terms of the left 
position operator $\hat{x}^L$ acted on the sliver: 
\beq
\forall \beta : \qquad \left( \exists \lambda \ {\rm such \ that} \ 
(\beta \cdot a^\dagger) \ket{\Xi} = (\lambda \cdot \hat{x}^L) \ket{\Xi} \right) 
\Leftrightarrow \rho_1 \beta = 0 \,.
\label{projdef}
\eeq
Let us show that (\ref{proj}) implies (\ref{projdef}). We will do that in two steps: 
First we prove that  (\ref{proj}) implies the ``$\Rightarrow$'' part of (\ref{projdef}). 
Indeed, assume that we can write $(\beta \cdot a^\dagger) \ket{\Xi}$ as 
$(\lambda \cdot \hat{x}^L) \ket{\Xi}$ for some $\lambda$, then:
\bea
(\lambda \cdot \hat{x}^L) \ket{\Xi} &=& \left( {i \over 2} \, 
\lambda \cdot \tilde{A}^+ E \, ( 1 + S) \, a^\dagger \right) \ket{\Xi}
\nonumber
\\
&=& \left( {i \over 2} \, (1 + S) \, E \tilde{A}^{+T} \lambda \right) 
\cdot a^\dagger \ket{\Xi}
\nonumber
\\
&=& (\beta \cdot a^\dagger) \ket{\Xi} \,,
\eea
where $\beta = {i \over 2} \, (1 + S) \, E \tilde{A}^{+T} \lambda$, and thus, 
from (\ref{proj}), $\rho_1 \beta = 0$.

\paragraph{}
Now let us prove that (\ref{proj}) implies the ``$\Leftarrow$'' part of (\ref{projdef}).
Remembering that $\tilde{A}^{+T} A^{+T} + \tilde{A}^{-T} A^{-T} = 1$, and assuming that 
$(1 + S)$ is invertible, we can write: 
\beq
\beta = ( 1 + S) \, E \, (\tilde{A}^{+T} A^{+T} + 
\tilde{A}^{-T} A^{-T}) \, \mu \,,
\label{betaequ}
\eeq
for some vector $\mu$. Now $\rho_1 \beta = 0$ implies (from (\ref{proj})) that 
$\rho_1 \, (1 + S) \, E \tilde{A}^{-T} A^{-T} \, \mu = 0$. Therefore
\beq
(1 + S) \, E \tilde{A}^{-T} A^{-T} \, \mu = (\rho_1 + \rho_2) \, 
(1 + S) \, E \tilde{A}^{-T} A^{-T} \, \mu = \rho_2 \, (1 + S) \, 
E \tilde{A}^{-T} A^{-T} \, \mu = 0 \,,
\eeq
where we have used (\ref{proj}) in the last step. Putting this into (\ref{betaequ}), we 
see that 
\beq
(\beta \cdot a^\dagger) \ket{\Xi} = \left( ( 1 + S) \, E \, \tilde{A}^{+T} A^{+T} \mu 
\cdot a^\dagger \right) \ket{\Xi} = {i \over 2} \, \left( 
( 1 + S) \, E \, \tilde{A}^{+T} \lambda \cdot a^\dagger \right) \ket{\Xi} = 
(\lambda \cdot \hat{x}^L) \ket{\Xi} \,,
\eeq
where we have defined $\lambda = -2 i \, A^{+T} \mu$. We thus have that (\ref{proj}) 
implies that $\rho_1$ is a left projector on the sliver in the sense of (\ref{projdef}).

\paragraph{}
We will now give a proof of (\ref{proj}) based on the factorization of the 
sliver and on the form of the star-product in the configuration basis (\ref{expect1}). 
For that, we 
will show that (\ref{proj}) is equivalent to the vanishing of 
\beq
\left( (\lambda \cdot \hat{x}^R) \ket{\Xi} \right) 
\star \ket{\Xi} \,,
\label{rightzero}
\eeq
where $\lambda$ is 
a given vector of numbers.

\paragraph{}
Indeed, using that $a \ket{\Xi} = - S \, a^{\dagger} \ket{\Xi}$, one gets
\bea
(\lambda \cdot \hat{x}^R) \ket{\Xi} &=& (\lambda \cdot \tilde{A}^- \hat{x}) 
\ket{\Xi} 
\nonumber \\
&=& \left( {i \over 2} \, \lambda \cdot \tilde{A}^- E \, (a - a^{\dagger}) 
\right) \ket{\Xi}
\nonumber \\
&=& \left( - {i \over 2} \, \lambda \cdot \tilde{A}^- E \, (1+S) \, a^{\dagger} 
\right) \ket{\Xi}
\nonumber \\
&=& \left( -{i \over 2} \, a^{\dagger} \cdot C \, (1+S) \, E \, \tilde{A}^{+T} 
\lambda \right) \ket{\Xi} \,,
\label{rightsliv}
\eea
where in the last line we used $\tilde{A}^+ = \tilde{A}^- C$.
Now recall that, in 
\cite{RSZhalf}, the authors proved that:
\beq
\left( e^{-a^{\dagger} \cdot C \beta_1} \ket{\Xi} \right) \star 
\left( e^{-a^{\dagger} \cdot C \beta_2} \ket{\Xi} \right) = 
e^{- {\cal C} (\beta_1, \, \beta_2)} \, e^{-a^{\dagger} \cdot C 
(\rho_1 \beta_1 + \rho_2 \beta_2)} \ket{\Xi} \,,
\label{slivproj1}
\eeq
where 
\beq
{\cal C} (\beta_1, \, \beta_2) = \half \, (\beta_1, \, \beta_2) \, 
{C \over (1+T) (1-X)} \, \left( \begin{array}{cc} 
X (1-T) \;&\; Z \\ Y \;&\; X (1-T) \end{array} \right) 
\left( \begin{array}{c} \beta_1 \\ \beta_2 \end{array} \right) \,,
\label{slivproj2}
\eeq
and $\beta_1$ and $\beta_2$ are vectors of numbers. Substituting $\beta_1 = 
t \, \beta$, where $t$ is a real number, and $\beta_2 = 0$ 
into (\ref{slivproj1}) and (\ref{slivproj2}), one 
gets
\beq
\left( e^{-a^{\dagger} \cdot C \beta \, t} \ket{\Xi} \right) \star \ket{\Xi} 
= e^{-\half \, t^2 \, \beta \cdot {C X (1-T) \over (1+T) (1-X)} \, \beta} \,  
e^{-a^{\dagger} \cdot C \rho_1 \beta \, t} \ket{\Xi} \,.
\eeq	
Differentiating this last equation with respect to $t$ and setting $t=0$ gives us
\beq
\left( (-a^{\dagger} \cdot C \beta) \ket{\Xi} \right) \star \ket{\Xi} = 
(-a^{\dagger} \cdot C \rho_1 \beta) \ket{\Xi} \,.
\label{linsliv} \eeq
Now we substitute $\beta \equiv {i \over 2} \, (1+S) \, E \, \tilde{A}^{+T} 
\lambda$ into (\ref{linsliv}) and we use (\ref{rightsliv}) to get
\beq
\left( (\lambda \cdot \hat{x}^R) \ket{\Xi} \right) \star \ket{\Xi} = 
\left( - {i \over 2} \, a^{\dagger} \cdot C \, \rho_1 \, (1+S) \, E \, 
\tilde{A}^{+T} \lambda \right) \ket{\Xi} \,.
\label{firstway}
\eeq
Therefore, (\ref{proj}) holds if and only if (\ref{rightzero}) vanishes.

\paragraph{}
We will now show that (\ref{rightzero}) indeed vanishes, by inserting two complete sets 
of zero-momentum states:
\bea
\left( (\lambda \cdot \hat{x}^R) \ket{\Xi} \right) \star \ket{\Xi}
&=&
\left( \int[d x_1] \ket{x_1} \bra{x_1} (\lambda \cdot \hat{x}^R) \ket{\Xi} \right) 
\star \left( \int[d x_2] \ket{x_2} \bra{x_2} \Xi \rangle \right)
\nonumber
\\
&=&
\left( \int[d x_1] \ket{x_1} \bra{x_1} (\hat{x} \cdot \tilde{A}^{-T} \lambda) 
\ket{\Xi} \right) 
\star \left( \int[d x_2] \ket{x_2} \bra{x_2} \Xi \rangle \right)
\nonumber
\\
&=& 
\int [d x_1] [d x_2] \left( x_1 \cdot (\tilde{A}^-)^T \lambda \right) \, \psi_\Xi 
(x_1^L, x_1^R) \, \psi_\Xi (x_2^L, x_2^R) \, \left( \ket{x_1} \star \ket{x_2} \right)
\nonumber
\\
&\sim& 
\int [d x_1] [d x_2] \left( x_1^R \cdot \lambda \right) \, \psi_\Xi 
(x_1^L, x_1^R) \, \psi_\Xi (x_2^L, x_2^R) \, \delta(x_1^R - x_2^L) \,
\ket{A^+ x_1^L + A^- x_2^R} \,,
\nonumber
\\
\label{x1r}
\eea
where $\psi_\Xi (x^L, x^R) = \bra{x} \Xi \rangle = \tilde{\cal N}^{26} \exp 
\left( -\half x^L \cdot 
K x^L \right) \,  \exp \left( -\half x^R \cdot K x^R \right)$ is the sliver 
wavefunction. We will now factor the integral over $x_1$ into an integral over 
the left-half string and an integral over the right-half string: 
$\int [d x_1] \sim \int [d x_1^L] [d x_1^R]$, and focus on the piece of the 
integral(\ref{x1r}) over $x_1^R$ only:
\beq
\int [d x_1^R] (x_1^R \cdot \lambda) \, e^{- x_1^R \cdot K x_1^R} = 0 \,,
\eeq
because the integrand is odd. Therefore we have that
\beq
\left( (\lambda \cdot \hat{x}^R) \ket{\Xi} \right) \star \ket{\Xi} = 0 \,.
\eeq
And thus, (\ref{proj}) indeed holds.

\subsection{Setting up the calculation}

Before calculating the star-product $\ket{x_1} \star \ket{x_2}$, 
we shall make more precise what we mean by {\em zero-momentum}: we drop all 
dependence of the zero-mode $x_0$ by setting 
$\left( x_{1,2} \right)_0 = 0$, and our vacuum $\ket{0}$ is amputated from its 
zero-mode part (we can thus normalize it such that $\bra{0} 0 \rangle = 1$).
Let us now express 
$\ket{x_1}$ and $\ket{x_2}$ in the oscillator basis, namely (\cite{RSZhalf}, \cite{GT}):
\beq
\ket{x_{1,2}} = K_0^{26} \exp \left(-x_{1,2} \cdot E^{-2} x_{1,2} - 2 i \, 
a^{\dagger} \cdot C E^{-1} x_{1,2} + 
\half \, a^{\dagger} \cdot a^{\dagger} \right) \ket{0} \,,
\eeq
where $K_0$ is a constant that can be calculated from 
the condition $\int [d x] \ket{x} \bra{x} = 1$, where the integration measure is 
defined as $[d x] = \prod_{n \geq 1}{d x_n}$. We get:   
\beq
K_0 = \det \left( \left({2 \over \pi}\right)^{1 \over 4} E^{-\half} \right) \,.
\eeq
We introduce now, for reasons that will become clear below, the following 
regularized states:
\beq
\ket{x_{1,2}, \eps} \equiv K_0^{26} \exp \left(-x_{1,2} \cdot E^{-2} x_{1,2} - 2 i \, 
a^{\dagger} \cdot C E^{-1} x_{1,2} + 
\half \, (1 - \eps) \, a^{\dagger} \cdot a^{\dagger} \right) \ket{0}.
\eeq
We obviously have $\ket{x_{1,2}} = \lim_{\eps \rightarrow 0} \ket{x_{1,2}, \eps}$, 
but we will take the limit only at the end of our calculation because we will 
encounter singularities at $\eps = 0$.

\paragraph{}
The star-product is calculated using the vertex $\ket{V_3}$ in the oscillator 
basis:
$$
\ket{x_1, \eps} \star \ket{x_2, \eps} = ~_1\langle x_1, \eps | 
~_2\langle x_2, \eps | V_3 \rangle_{123}, 
$$
where
$$
\ket{V_3} = \exp \left( -\half \sum_{{r,s}\atop m, n \geq 1} 
a_m^{(r) \dagger} \cdot V_{mn}^{rs} a_n^{(s) \dagger} \right) \ket{0}_{123}.
$$
A summation over Lorentz indices is implicit in the $\cdot$ product. Using the 
following formula (\cite{KP}, \cite{RSZclassical}):
\bea 
&& \langle 0 | \exp\Big(\lambda_i a_i -{1\over 2} P_{ij} a_i a_j\Big)
\exp\Big(\mu_i a^\dagger_i -{1\over 2} Q_{ij} a^\dagger_i
a^\dagger_j\Big)
|0\rangle
\nonumber \\
=&& \det(K)^{-1/2} \exp\Big(\mu^T\, K^{-1} \lambda -{1\over
2}
\lambda^T \,Q \, K^{-1} \lambda - {1\over 2}\mu^T\, K^{-1}P\mu\Big)\,
,\,\,
K\equiv 1- PQ \ ,
\label{formula}
\eea
and setting $P = -(1- \eps) \left( \begin{array}{cc} 1 & 0 \\ 0 & 1 \end{array}
\right)$, $Q = \Nu \equiv \left( \begin{array}{cc} V^{11} & V^{12} \\ 
V^{21} & V^{22} \end{array}\right)$, $\lambda = \left( \begin{array}{c} 
2 i E^{-1} x_1 \\ 2 i E^{-1} x_2  \end{array} \right)$ and $\mu = -\left( 
\begin{array}{c} V^{13} \, a^{(3)\dagger} \\ V^{23} \, 
a^{(3)\dagger}\end{array}\right)$,
we get:
\bea
&& \ket{x_1, \eps} \star \ket{x_2, \eps} = 
\nonumber 
\\
&& K_0^{52} \left\{ \det \left(
1+(1 - \eps) \, \Nu \right)^{-1/2}\right\}^{26} 
\exp \left( -x_1\cdot E^{-2} x_1 - x_2\cdot E^{-2} x_2 \right) \times
\nonumber \\ 
&& \times \exp \left\{ \mu^T \cdot
{1 \over 1+(1 - \eps) \, \Nu} \, \lambda -\half \lambda^T \cdot  
{\Nu \over 1 + (1 - \eps) \, \Nu} \, \lambda + 
\half \mu^T \cdot {1 - \eps \over 1+(1 - \eps) \, \Nu} \, \mu - 
\half a^{(3)\dagger} \cdot V^{33} \, a^{(3)\dagger} \right\} \ket{0}.
\nonumber
\\
\label{start} 
\eea
The inverse of the matrix $(1+(1 - \eps) \, \Nu) = 
\left( \begin{array}{cc} 1 + (1 - \eps) \, C X & (1 - \eps) \, C Y \\ 
(1 - \eps) \, C Z & 1 + (1 - \eps) \, C X \end{array} \right)$ can be found as 
follows\footnote{Note that the submatrices in this block-matrix do not commute 
with each other, making the task of finding the inverse nontrivial.}: From 
the inverse calculated numerically, we infer the following Ansatz for the inverse: 
$\left( \begin{array}{cc} \mu + W & - W \, C \\ - C \, W & \mu + C \, W \, C 
\end{array} \right)$, where $\mu$ is a real number multiplying the identity matrix 
and $W$ is an unknown matrix. To find $W$, we ask that the following matrix $M$ be 
block-diagonal:
\beq
M = \left( \begin{array}{cc} M_{11} \;&\; M_{12} \crbig M_{21} \;&\; M_{22} 
\end{array} \right) \equiv \left( \begin{array}{cc} \mu + W \;&\; - W \, C \crbig 
- C \, W \;&\; \mu + C \, W \, C \end{array} \right) \cdot \left( \begin{array}{cc} 
1 + (1 - \eps) \, C X \;&\; (1 - \eps) \, C Y \crbig 
(1 - \eps) \, C Z \;&\; 1 + (1 - \eps) \, C X \end{array} \right) \,.
\label{preinverse}
\eeq
Requiring that $M_{12} = M_{21} = 0$, we get $W = \mu \, Z \left( 
{1 \over 1 - \eps} + C \, X - Z \right)^{-1}$. Now substituting this expression 
for $W$ into (\ref{preinverse}), we find 
$M_{11} = \mu \left( 1 + (1 - \eps) \, C \, X + (1 - \eps) \, Z \right)$ and 
$M_{22} = C \, M_{11} \, C$. The inverse is therefore 
$\left( \begin{array}{cc} M_{11}^{-1} & 0 \\ 0 & M_{22}^{-1} \end{array} \right) 
\cdot \left( \begin{array}{cc} \mu + W & - W \, C \\ - C \, W & \mu + C \, W \, C 
\end{array} \right)$. Namely: 
\beq
(1+(1 - \eps) \, \Nu)^{-1} = {1 \over 1 - \eps} \, 
\left( \begin{array}{cc} 
H^+(\eps) \, ({1 \over 1 - \eps} + C X) \,  
H^-(\eps) \; & \; -H^+(\eps) \, Z \, H^-(\eps) \, C  
\crbig 
-C \, H^+(\eps) \, Z \, H^-(\eps) \; & \; 
C \, H^+(\eps) \, ({1 \over 1 - \eps} + C X) \, H^-(\eps) \, C
\nonumber \end{array} \right) \,,
\label{inverse}
\eeq
where we have defined 
\beq
H^\pm(\eps) \equiv \left( {1 \over 1 - \eps} + C X \pm Z \right)^{-1} \,.
\eeq
The expression (\ref{inverse}) was constructed by requiring that it be a left inverse of 
$(1+(1 - \eps) \, \Nu)$, but one can check that this is also a right inverse.

\paragraph{}
The matrix $H^+(\eps)$ can be expressed more explicitly in terms of 
the inverse of $({1 \over 1-\eps} - X)$, which behaves well because it commutes 
with $C$, $X$, 
$Y$ and $Z$. In order to find the form of this inverse, we make the following Ansatz: 
$H^+(\eps) = f(X) \, (\alpha - \beta \, C \, X + \gamma \, Y)$, where $\alpha$, 
$\beta$ and $\gamma$ are real numbers, and $f(X)$ is a function involving only 
the matrix $X$ (and the identity matrix). One can solve for all these parameters and 
we find:
\bea
H^+(\eps) 
&=& {1 \over 2 - \eps} \, {1 \over {1 \over 1 - \eps} - X} 
\left( 1 - (1 - \eps) \, C X + (1 - \eps) \, Y \right) 
\nonumber \\
&=& \half \, {1 \over 1-X} \left( 1 - C X+Y \right) + \Or(\eps)
\label{posinv} \\
&=& {1 \over 4} \left( \begin{array}{cc} (1+U_{oo})^{-1}(5-U_{oo}) \;&\; 
-\sqrt 3 i \, (1+U_{oo})^{-1} U_{oe} \crbig
\sqrt 3 i \, (1-U_{ee})^{-1} U_{eo} \;&\; 3 \end{array} \right) + \Or(\eps).
\label{posinvpar}
\eea
In the second line, we wrote only the first term in the power expansion in $\eps$, 
and in the third line we expressed the above line in the parity notation. 
In the rest of the paper, we will often neglect to write the ${\cal O}(\eps)$ 
terms when they are not relevant.

\paragraph{}
The matrix $H^-(\eps)$ is a little bit more problematic. It can be seen to be 
$$
H^-(\eps) = 
- {1 \over \eps} \, {1 \over {1 \over 1 - \eps} + X} 
\left( -1 + (1 - \eps) \, C X + (1 - \eps) \, Y \right) \,.
$$
We see that it is singular when $\eps = 0$; this is precisely why we have to use 
the regulator $\eps$.  
Let us define the singular and regular parts of $H^-(\eps)$ in the following way:
$$
H^-(\eps) \equiv {1 \over \eps} \, H^-_{\rm sing} + H^-_{\rm reg} + \Or(\eps) \,,
$$
where
\bea
H^-_{\rm sing} &=& {1 \over 1 + X} \, (1 - C X - Y)
\nonumber \\
&=& {1 \over 1+X} \, \left( \begin{array}{cc} 
1 - U_{oo} \;&\; {\sqrt{3} \over 3} i \, U_{oe} \crbig
- {\sqrt{3} \over 3} i \, U_{eo} \;&\; {1 \over 3} - {1 \over 3} \, U_{ee}
\end{array} \right) \,,
\nonumber \\
H^-_{\rm reg}
&=& {1 \over (1+X)^2} \, (-1+2 Y+X Y+C(2 X+X^2))
\nonumber \\
&=& {1 \over (1+X)^2} \, \left( \begin{array}{cc} 
-1+{7 \over 3} \, U_{oo} - {2 \over 3} \, U_{oo}^2 \;&\; 
-{7 \over 9} \, \sqrt{3} i \, U_{oe} + {2 \over 9} \, \sqrt{3} i \, U_{oo}U_{oe}
\crbig
{7 \over 9} \, \sqrt{3} i \, U_{eo} + {2 \over 9} \, \sqrt{3} i \, U_{ee}U_{eo} 
\;&\; {5 \over 9} + {11 \over 9} \, U_{ee} + {2 \over 9} \, U_{ee}^2 
\end{array} \right) \,.
\nonumber
\eea

\paragraph{}
With these expressions in hand, one can now calculate explicitly 
the star-product (\ref{start}).

\section{Technical details of the calculation} \label{s3}

To begin, let us write 
\beq
\ket{x_1, \eps} \star \ket{x_2, \eps} = 
K_0^{52} \left( \det(1 + (1 - \eps) \Nu)^{-\half} \right)^{26} \, 
\Omega(x_1, x_2) \, e^{\ell(a^\dagger, x_1, x_2)} \,
e^{\half a^{\dagger}\cdot \Gamma a^{\dagger}} \ket{0} \,,
\label{descr}
\eeq
where $\Omega$ is the part of (\ref{start}) quadratic in $x$, $e^\ell$ is the part
linear in $x$, and $\Gamma$ doesn't depend on $x$.

\paragraph{}
We will start by calculating $\Gamma$ in subsection \ref{sub1}, in the next subsection 
we will calculate the terms proportional to ${1 \over \eps}$ in $\Omega$, 
leaving the regular terms in $\Omega$ 
for subsection \ref{sub3}. Then in subsection \ref{sub4}, we will calculate the linear 
term $\ell$. And finally we will look at the overall normalization in subsection 
\ref{sub5}, and we will derive the form of the star-product in the wavefunction 
representation.

\subsection{Term quadratic in $a^{\dagger}$} \label{sub1}
 Our goal here is to prove that 
$\Gamma = 1$.

\paragraph{}
From (\ref{start}), we can write 
\bea
\Gamma &=& (V^{31} \;\; V^{32}) \, {1 - \eps \over 1 + (1 - \eps) \, \Nu} \left( 
\begin{array}{c} V^{13} \crbig V^{23} \end{array} \right) - V^{33}
\nonumber \\
&=& (C Y \;\; C Z) \, {1 - \eps \over 1+(1 - \eps) \, \Nu} \left( \begin{array}{c} 
CZ \crbig CY \end{array} \right) - CX
\nonumber \\
&=& \left( \Lambda + C \Lambda C \right) - C X \, ,
\label{Gamma1} \eea
where 
\beq
\Lambda \equiv Y \, H^+(\eps) \, \left({1 \over 1 - \eps} + C X 
\right) \, H^-(\eps) \, Y - Y \, H^+(\eps) \, Z \,  H^-(\eps) \, Y \, C \,.
\label{Gamma2} 
\eeq
From (\ref{Gamma1}), it is clear that $C \Gamma C = \Gamma$, which implies that 
$\Gamma_{oe} = \Gamma_{eo} = 0$; and also $\Gamma_{oo} = 2 \, \Lambda_{oo} + X_{oo}$ 
and $\Gamma_{ee} = 2 \, \Lambda_{ee} - X_{ee}$. From (\ref{Gamma2}), we see that
\bea
\Lambda_{oo} &=& \left\{ Y \left({1 \over 1 - \eps}+C X + Z \right)^{-1} 
\left({1 \over 1 - \eps}+C X + Z \right) \left({1 \over 1 - \eps}+C X - Z \right)^{-1} Y 
\right\}_{oo} 
\nonumber \\
&=& \left\{ Y \, H^-(\eps) \, Y \right\}_{oo} \,,
\nonumber \crbig
\Lambda_{ee} &=& \left\{ Y \left({1 \over 1 - \eps} + C X + Z \right)^{-1} 
\left({1 \over 1 - \eps}+C X-Z \right) 
\left({1 \over 1 - \eps}+C X-Z \right)^{-1} Y \right\}_{ee} 
\nonumber \\
&=& \left\{ Y \, H^+(\eps) \, Y \right\}_{ee} \,.
\eea
We will now proceed further by using the parity notation. Let us first calculate 
$\Lambda_{ee}$ (denoting by ``$\cdots$'' matrix elements that we don't need to know):
\bea
\Lambda_{ee} &=& \left\{ Y \, H^+(\eps) \, Y \right\}_{ee}
\nonumber \\
&=& \left\{ {1 \over 3 \cdot 4} \, \left( \begin{array}{cc} 1+U_{oo} \; &\;  
-\sqrt{3}i \, U_{oe} \crbig
\sqrt{3}i \, U_{eo} \; & \; 1-U_{ee} \end{array} \right)
\left( \begin{array}{cc} (1+U_{oo})^{-1} (5-U_{oo}) \; & \; 
-\sqrt{3}i (1+U_{oo})^{-1} U_{oe} \crbig
\sqrt{3}i (1-U_{ee})^{-1} U_{eo} \; & \; 3 \end{array} \right)Y \right\}_{ee}
\nonumber \\
&=& \left\{ {1 \over 3 \cdot 4 \cdot 3} \left( \begin{array}{cc} \cdots \;&\; \cdots \crbig
6\sqrt{3}i \, U_{eo} (1+U_{oo})^{-1} \;&\; 6 \end{array} \right) 
\left( \begin{array}{cc} \cdots \; &\;  
-\sqrt{3}i \, U_{oe} \crbig \cdots \; & \; 1-U_{ee} \end{array} \right)
\right\}_{ee}
\nonumber \\
&=& {1 \over 3} \, (2+U_{ee}) \,,
\nonumber
\eea
and thus 
$$
\Gamma_{ee} = {2 \over 3} \, (2+U_{ee}) - X_{ee} = {2 \over 3} \, (2+U_{ee}) - 
{1 \over 3} \, (1+2 U_{ee}) = 1 \,.
$$
Let us now turn to $\Lambda_{oo}$. Its singular part is
\bea
(\Gamma_{oo})_{\rm sing}
&=& \left\{ Y \, H^-_{\rm sing} \, Y \right\}_{oo}
\nonumber \\
&=&
\left\{ {1 \over 3 \, (1 + X)} \, \left( \begin{array}{cc} 1+U_{oo} \; &\;  
-\sqrt{3}i \, U_{oe} \crbig
\sqrt{3}i \, U_{eo} \; & \; 1-U_{ee} \end{array} \right)
\left( \begin{array}{cc} 
1 - U_{oo} \;&\; {\sqrt{3} \over 3} i \, U_{oe} \crbig
-{\sqrt{3} \over 3} i \, U_{eo} \;&\; {1 \over 3} - {1 \over 3} \, U_{ee}
\end{array} \right) Y \right\}_{oo}
\nonumber \\
&=& \left\{ {1 \over 3 \, (1 + X)} \, \left( \begin{array}{cc} 0 \;&\; 0 \crbig 
\cdots \;&\; \cdots \end{array} \right) Y \right\}_{oo} = 0 \,,
\eea
and its regular part is
\bea
&& (\Gamma_{oo})_{\rm reg} = 
\left\{ Y \, H^-_{\rm reg} \, Y \right\}_{oo} =
\nonumber \\ 
&=&
\left\{ {1 \over 3\, (1+X)^2} \left( \begin{array}{cc} 1+U_{oo} \; &\;  
-\sqrt{3}i \, U_{oe} \crbig
\sqrt{3}i \, U_{eo} & 1-U_{ee} \end{array} \right)
\left( \begin{array}{cc} 
-1+{7 \over 3} \, U_{oo} - {2 \over 3} \, U_{oo}^2 & 
-{7 \over 9} \, \sqrt{3} i \, U_{oe} + {2 \over 9} \, \sqrt{3} i \, U_{oo}U_{oe}
\crbig
{7 \over 9} \, \sqrt{3} i \, U_{eo} + {2 \over 9} \, \sqrt{3} i \, U_{ee}U_{eo} 
\;&\; {5 \over 9} + {11 \over 9} \, U_{ee} + {2 \over 9} \, U_{ee}^2 
\end{array} \right)
 Y \right\}_{oo}
\nonumber \\
&=&
\left\{ {1 \over 27\, (1+X)^2} \, \left( \begin{array}{cc} 4 + 2 \, U_{oo} - 2 U_{oo}^2 \;&\; 
-4 \sqrt{3} i \, U_{oe} + 2 \sqrt{3} i \, U_{oo} U_{oe} \crbig
\cdots \;&\; \cdots \end{array} \right)
\left( \begin{array}{cc} 1+U_{oo} \;&\; \cdots \crbig \sqrt{3} i \, U_{eo} \;&\; 
\cdots \end{array} \right) \right\}_{oo}
\nonumber \\
&=& {1 \over 27} \, \left\{ {1 \over (1+X)^2} \right\}_{oo} 
(16 - 12\, U_{oo}^2 + 4\, U_{oo}^3)
\nonumber \\
&=& {1 \over 3} + {1 \over 3} \, U_{oo} \,.
\nonumber
\eea
Thus $\Gamma_{oo} = {2 \over 3} + {2 \over 3} \, U_{oo} + \left( {1 \over 3} - 
{2 \over 3} \, U_{oo} \right) = 1$,
which concludes the proof that $\Gamma = 1$.

\subsection{The delta function} \label{sub2}
Now that the reader is used to the formalism, we can track down the delta function 
that we expect. We will see that it is of the form
$$
\lim_{{\eps \rightarrow 0} \atop \eps > 0} 
\left( \det \left({1 \over \pi \, \eps} \, M \right)^\half 
\, e^{-{1 \over \eps} \, 
y \cdot M y} \right) = \delta (y) \,,
$$
where $M$ is strictly positive definite.
For this, we will concentrate on the terms proportional to ${1 \over \eps}$ 
appearing in the exponential of (\ref{start}). We have already shown that such terms 
don't appear in the term quadratic in $a^\dagger$ and we will show later that they neither 
appear in the term linear in $a^\dagger$.

\paragraph{}

\noindent For now let us calculate the term quadratic in $x$.
From (\ref{start}) this is 
\bea
\Omega &\equiv& \exp \left\{ -x_1 \cdot E^{-2} x_1 - x_2 \cdot E^{-2} x_2 + 2 \, 
(x_1 \cdot E^{-1} \ , \  
x_2 \cdot E^{-1}) \, \left({\Nu \over 1 + (1 - \eps) \Nu}\right)
\left( \begin{array}{c} E^{-1}x_1 \crbig 
E^{-1}x_2 \end{array} \right) \right\}
\nonumber \\
&=& \exp \left\{ (x_1 \cdot E^{-1} \ , \  
x_2 \cdot E^{-1}) \, W \left( \begin{array}{c} E^{-1}x_1 \crbig 
E^{-1}x_2 \end{array} \right) \right\} \,,
\nonumber
\eea
where 
\beq
W \equiv (1 + 2 \, \eps) \left(1 - 2 \, {1 - \eps \over 1 + (1 - \eps) \Nu} \right) + 
{\cal O}(\eps) \,.
\label{W} \eeq
Again, we separate $\Omega$ and $W$ into a regular and a singular part:
$\Omega = \Omega_{\rm reg} \, \Omega_{\rm sing}$, 
$W = W_{\rm reg} + {1 \over \eps} \, W_{\rm sing}$, with
\bea
\Omega_{\rm reg} &=& \exp \left\{ (x_1 \cdot E^{-1} \;,\; 
x_2 \cdot E^{-1}) \, W_{\rm reg} \left( \begin{array}{c} E^{-1}x_1 \crbig 
E^{-1}x_2 \end{array} \right) \right\} \,,
\label{Omegareg}
\\
\Omega_{\rm sing} &=& \exp \left\{ (x_1 \cdot E^{-1} \;,\; 
x_2 \cdot E^{-1}) \, {1 \over \eps} \, W_{\rm sing} \left( 
\begin{array}{c} E^{-1}x_1 \crbig 
E^{-1}x_2 \end{array} \right) \right\} \,.
\label{Omegasing}
\eea

\paragraph{}
We will come back later to the regular part, and we focus now on the singular 
piece
\bea
\hspace{-36pt} W_{\rm sing} &=& \left( \begin{array}{cc} \left(W_{\rm sing} \right)_{11} 
\;&\; \left( W_{\rm sing} \right)_{12} \crbig 
C \left(W_{\rm sing} \right)_{12} C \;&\; C \left(W_{\rm sing} \right)_{11} C 
\end{array} \right) \,,
\nonumber
\\
\hspace{-36pt} \left(W_{\rm sing} \right)_{11} 
&=& -2 \, \left[ H^+(\eps) \, \left( {1 \over 1 - \eps} + C X \right) \, H^-_{\rm sing} 
\right]_{\eps = 0} \,,
\quad
\left(W_{\rm sing} \right)_{12} = 2 \, \left[ H^+(\eps) \, Z \, H^-_{\rm sing} 
\right]_{\eps = 0} \, C 
\label{Wsing}
\eea
In order to calculate $\Omega_{\rm sing}$, we will need the following identities: 
\bea
H^+(0) \, Z \, H^-_{\rm sing} &=& \half H^-_{\rm sing} \,,
\label{br1}
\crbig 
H^+(0) \, \left(1 + C X \right) \, H^-_{\rm sing} 
&=& \half H^-_{\rm sing} \,,
\label{br2}
\crbig
E^{-1} A^+ \tilde{A^+} E &=& \half \, \left( \begin{array}{cc} 1 \;&\; 
E^{-1} \left( \Xgt_{eo} \right)^{-1_L} E \crbig 
E^{-1} \Xgt_{eo} E \;&\; 1 \end{array} \right) 
\nonumber
\\
&=& 
\half \, \left( \begin{array}{cc} 
1 \;&\; -{i \over \sqrt{3}} \, (1 - U_{oo})^{-1} U_{oe} \crbig
\sqrt{3} i \, (1 - U_{ee})^{-1} U_{eo} \;&\; 1 \end{array} \right) \,,
\label{EAAE}
\crbig 
H^-_{\rm sing} \, E^{-1} A^+ &=& 0
\label{propsing} \,.
\eea
The first one can be shown straightforwardly in the parity notation, the second one 
follows from the first one, and the third one can be established using the definitions 
of $A^\pm$, $\tilde{A}^\pm$ and $\Xgt$ as well as (\ref{one}) and 
(\ref{two}); $\left( \Xgt_{eo} \right)^{-1_L}$ stands for the {\em left} inverse 
of $\Xgt_{eo}$. Finally, to show the last identity, one can show in the parity notation, 
using (\ref{EAAE}), that $H^-_{\rm sing} \, E^{-1} A^+ \tilde{A}^+ E = 0$. 
The identity then follows because $\tilde{A}^+$ is invertible on the right.

\paragraph{}
Let us now calculate the term, in the exponential of $\Omega_{\rm sing}$, 
quadratic in $x_1$:
\bea
x_1 \cdot E^{-1} \left(W_{\rm sing} \right)_{11} E^{-1} x_1 
&=& -2 \, x_1 \cdot E^{-1} \, H^+(0) \, \left(1 + C X \right) \, 
H^-_{\rm sing} \, E^{-1} x_1
\nonumber \\
&=& - x_1 \cdot E^{-1} \, H^-_{\rm sing} \, E^{-1} x_1
\nonumber \\
&=& - x_1 \cdot E^{-1} \, H^-_{\rm sing} \, E^{-1} \left( A^+ x_1^L + 
A^- x_1^R \right)
\nonumber \\
&=& - x_1 \cdot E^{-1} \, H^-_{\rm sing} \, E^{-1} A^- x_1^R 
\nonumber \\
&=& - x_1^R \cdot (A^-)^T E^{-1} \, H^-_{\rm sing} \, E^{-1} 
\left( A^+ x_1^L + A^- x_1^R \right)
\nonumber \\
&=& - x_1^R \cdot (A^-)^T E^{-1} \, H^-_{\rm sing} \, E^{-1} A^- x_1^R
\nonumber \\
&=& - x_1^R \cdot M x_1^R \,,
\nonumber
\eea
where
\beq
M \equiv (A^-)^T E^{-1} \, H^-_{\rm sing} \, E^{-1} A^- 
\label{M} \,.
\eeq 
The term quadratic in $x_2$ is calculated similarly:
\bea
x_2 \cdot E^{-1} \left(W_{\rm sing} \right)_{22} E^{-1} x_2 
&=& - x_2 \cdot E^{-1} C \, H^-_{\rm sing} \, C E^{-1} x_2
\nonumber \\
&=& - x_2 \cdot E^{-1} C \, H^-_{\rm sing} \, E^{-1} C \left( A^+ x_2^L + 
A^- x_2^R \right)
\nonumber \\
&=& - x_2 \cdot E^{-1} C \, H^-_{\rm sing} \, E^{-1} \left( A^- x_2^L + 
A^+ x_2^R \right)
\nonumber \\
&=& - x_2 \cdot C E^{-1} \, H^-_{\rm sing} \, E^{-1} A^- x_2^L 
\nonumber \\
&=& - x_2^L \cdot M x_2^L \,.
\nonumber
\eea
The calculation of the terms containing $x_1$ and $x_2$ is similar 
to the above ones and we can see that $\Omega_{\rm sing}$ can be written
\beq
\Omega_{\rm sing} = 
\exp \left( -{1 \over \eps} \, (x_1^R - x_2^L) \cdot M (x_1^R - x_2^L) \right) 
\,.
\label{predelta}
\eeq

\paragraph{}
We can see that $M$ is positive definite by writing 
$M = B^T B$, where
\beq
B = {1 \over \sqrt{2}} \, (1 + X)^{-1/2} \, (1 + C) \, (1 - Y)^{1/2} 
E^{-1} \, A^- \,.
\eeq
To verify this, the reader should keep in mind that $(X^2)^{1/2} = -X$ if the square 
root is defined to have only positive eigenvalues. Indeed we assumed that the spectrum 
of $X$ is contained in the interval $[-{1 \over 3}, 0]$. 
Though there is still no proof of this, there is 
compelling numerical evidence that it is indeed true\footnote{I thank B.~Zwiebach for 
a discussion of this point}.

\paragraph{}
Unfortunately, $M$ doesn't seem to be strictly positive definite since 
$(1 + C)$ is singular; we thus expect $B$, and therefore $M$, to have some zero 
eigenvalues. Numerical analysis shows that this might not be the case: If we 
truncate all matrices to a size $N$, we find indeed $N/2$ ``small'' eigenvalues, 
but the corresponding normalized eigenvectors all have some ``non-small'' components 
which have indices $\geq N/2$. Therefore we believe that, 
when we take the limit $N \rightarrow \infty$, these eigenvectors cannot be normalized 
anymore, and thus disappear from the spectrum. 
A better understanding of this phenomenon would be interesting to 
pursue.

\paragraph{}
For now, we will assume that $M$ is strictly positive definite and thus, if we 
restrict $\eps$ to be positive, (\ref{predelta}) gives, in the limit $\eps 
\rightarrow 0$:
\beq
\Omega_{\rm sing} = \det \left( {1 \over \pi \, \eps} \, M \right)^{-\half} 
\delta (x_1^R - x_2^L) \,,
\eeq
which was expected if we think of the star-product to glue the right-half of 
the first string to the left-half of the second string.

\subsection{Regular quadratic term in $x$} \label{sub3}
We now wish to calculate $\Omega_{\rm reg}$ (\ref{Omegareg}). For this, we first 
note that $W_{\rm reg}$ gets contributions from two different kinds of terms: We will 
write $W_{\rm reg} = W_{\rm reg}^{(1)} + W_{\rm reg}^{(2)}$, where $W_{\rm reg}^{(1)}$ 
denotes the terms that are a product of a term proportional to $\eps$ and a singular term 
(proportional to ${1 \over \eps}$) coming from the singular part of 
$H^-(\eps)$; $W_{\rm reg}^{(2)}$ denotes 
all the other terms.

\paragraph{}
Let us first calculate the contribution from $W_{\rm reg}^{(1)}$ 
in $\Omega_{\rm reg}$. As defined above, $W_{\rm reg}^{(1)}$ can be expressed from 
(\ref{Omegasing}) and (\ref{Wsing}) by keeping the terms proportional to $\eps$ that 
are multiplying 
$H^-_{\rm sing}$. These are: 
\bea
W_{\rm reg}^{(1)} &=& \left( \begin{array}{cc} \left( W_{\rm reg}^{(1)} 
\right)_{11} \;&\; \left( W_{\rm reg}^{(1)} \right)_{12} \crbig 
C \left( W_{\rm reg}^{(1)} \right)_{12} C \;&\; 
C \left( W_{\rm reg}^{(1)} \right)_{11} C \end{array} \right) \,,
\nonumber
\crbig
\left(W_{\rm reg}^{(1)} \right)_{11} &=& 2 \, \left( W_{\rm sing} \right)_{11} 
-2 \, \left[ {d \over d \eps} \, \left( 
H^+(\eps) \, \left( {1 \over 1 - \eps} + C X \right) \right) \right]_{\eps = 0}
H^-_{\rm sing} + \Or(\eps)
\nonumber 
\\
&=& 2 \, \left( W_{\rm sing} \right)_{11} 
-2 \, H^+(0)^2 \, Z \, H^-_{\rm sing} + {\cal O}(\eps)
\nonumber
\\
&=& 2 \, \left( W_{\rm sing} \right)_{11} 
- H^+(0) \, H^-_{\rm sing} + {\cal O}(\eps) \,,
\nonumber
\crbig
\left( W_{\rm reg}^{(1)} \right)_{12} &=& 2 \, \left( W_{\rm sing} \right)_{12}
+ 2 \, \left[ {d \over d \eps} \, \left( H^+(\eps) \, Z \right) \right]_{\eps = 0} \, 
H^-_{\rm sing} C + \Or(\eps)
\nonumber
\\
&=& 2 \, \left( W_{\rm sing} \right)_{12} 
-H^+(0) \, H^-_{\rm sing} C + {\cal O}(\eps) \,,
\label{Wreg1}
\eea
where we have used (\ref{br1}). After substituting $x = A^+ x^L + 
A^- x^R$ in (\ref{Omegareg}) and using $x_1^R = x_2^L$, it is straightforward to see 
that the $W_{\rm sing}$ terms in (\ref{Wreg1}) will disappear, leaving us with 
\bea
&& (x_1 \cdot E^{-1} \;,\; x_2 \cdot E^{-1}) \, W_{\rm reg}^{(1)} \left( 
\begin{array}{c} E^{-1}x_1 \crbig E^{-1}x_2 \end{array} \right) = 
\nonumber
\\
&& = - (x_1 \cdot E^{-1} \;,\; x_2 \cdot E^{-1}) \left( \begin{array}{cc} 
H^+(0) H^-_{\rm sing} \;&\; H^+(0) H^-_{\rm sing} C \crbig
C H^+(0) H^-_{\rm sing} \;&\; C H^+(0) H^-_{\rm sing} C \end{array} \right) 
\left( \begin{array}{c} E^{-1} A^+ \, x_1^L + E^{-1} A^- x_2^L \crbig 
E^{-1} A^+ \, x_2^L + E^{-1} A^- x_2^R \end{array} \right) 
\nonumber
\\
&& = - 2 \, (x_1^L \cdot A^{+T} E^{-1} + x_2^L \cdot A^{-T} E^{-1} \;,\; 
x_2^L \cdot A^{+T} E^{-1} + x_2^R \cdot A^{-T} E^{-1}) 
\left( \begin{array}{c} H^+(0) H^-_{\rm sing} E^{-1} A^- \, x_2^L \crbig 
C H^+(0) H^-_{\rm sing} E^{-1} A^- \, x_2^L \end{array} \right) \,,
\nonumber
\\
\label{dechet1}
\eea
where we have made use of (\ref{propsing}).

\paragraph{}
We now turn to $W_{\rm reg}^{(2)}$.  We have:
\bea
W_{\rm reg}^{(2)} &=& \left( \begin{array}{cc} \left( W_{\rm reg}^{(2)} \right)_{11} 
\;&\; \left( W_{\rm reg}^{(2)} \right)_{12} \crbig
C \left( W_{\rm reg}^{(2)} \right)_{12} C 
\;&\; C \left( W_{\rm reg}^{(2)} \right)_{11} C \end{array} \right) \,,
\\
\left( W_{\rm reg}^{(2)} \right)_{11} &=& 1 - 2 \, H^+(0) \, \left( 1 + C X \right) \, 
H^-_{\rm reg} \,,
\\
\left( W_{\rm reg}^{(2)} \right)_{12} &=& 2 \, H^+(0) \, Z \, H^-_{\rm reg} \, C \,.
\eea
We will need the following identities:
\bea
&& \hspace{-24pt} 
A^{+T} E^{-1} \left( W_{\rm reg}^{(2)} \right)_{11} E^{-1} A^+ 
= - A^{+T} E^{-2} A^+ \,,
\label{sh1}
\\
&& \hspace{-24pt} 
A^{+T} E^{-1}  \left( \left( W_{\rm reg}^{(2)}\right)_{12} + C \, 
\left( W_{\rm reg}^{(2)} \right)_{12}^T \, C \right) 
E^{-1} A^- 
= -2 \, A^{+T} E^{-2} A^- \,,
\label{sh2}
\\
&& \hspace{-24pt} 
H^+(0) \, E^{-1} A^- = \half 
\, E^{-1} A^- \,,
\label{sh3}
\\
&& \hspace{-24pt}
A^{-T} E^{-1} H^+(0) H^-_{\rm sing} E^{-1} A^- = A^{-T} E^{-1} \left( \half - 
H^+(0) \, (1 + C X - Z) \, H^-_{\rm reg} \right) E^{-1} A^- \,.
\label{sh4}
\eea
Equations (\ref{sh1}) - (\ref{sh3}) can be established straightforwardly 
with the parity notation 
(Note that to use the parity notation, one must first multiply (\ref{sh1}) on the left 
by $E \tilde{A}^{+T}$ and on the right by $\tilde{A}^+ E$, and also multiply (\ref{sh2}) 
on the left by $E \tilde{A}^{+T}$ and on the right by $\tilde{A}^- E$. This is allowed 
since $\tilde{A}^{\pm}$ have a right inverse and thus $\tilde{A}^{\pm T}$ have a left 
inverse).
Equation (\ref{sh4}) can be proved using $(1 + C X - Z) \, H^-_{\rm reg} = 
1 - H^-_{\rm sing}$, which can be shown from the expressions of $H^-_{\rm reg}$ and 
$H^-_{\rm sing}$.
Substituting $x = A^+ x^L + A^- x^R$ and $x_1^R = x_2^L$, we get
\bea
&& (x_1 \cdot E^{-1} \;,\; x_2 \cdot E^{-1}) \, W_{\rm reg}^{(2)} \left( 
\begin{array}{c} E^{-1}x_1 \crbig E^{-1}x_2 \end{array} \right) = 
\nonumber
\\
&& = (x_1^L \cdot A^{+T} + x_2^L \cdot A^{-T} \,, \ 
x_2^L \cdot A^{+T} + x_2^R \cdot A^{-T}) E^{-1} \, W_{\rm reg}^{(2)} \, E^{-1} 
\left( \begin{array}{c} A^+ x_1^L + A^- x_2^L \crbig A^+ x_2^L + A^- x_2^R 
\end{array} \right) \,.
\label{dechet2}
\eea
Using (\ref{sh1}) - (\ref{sh4}), it is straightforward to calculate each term in 
(\ref{dechet1}) and (\ref{dechet2}). And as a final result, we get, as was expected:
\bea
\Omega_{\rm reg} &=& \exp \left(  (x_1 \cdot E^{-1} \;,\; x_2 \cdot E^{-1}) \, 
W_{\rm reg}^{(1)} \left( \begin{array}{c} E^{-1}x_1 \crbig E^{-1}x_2 \end{array} 
\right) + (x_1 \cdot E^{-1} \;,\; x_2 \cdot E^{-1}) \, 
W_{\rm reg}^{(2)} \left( \begin{array}{c} E^{-1}x_1 \crbig E^{-1}x_2 \end{array} 
\right) \right)
\nonumber
\\
&=& \exp \left( - (A^+ x_1^L + A^- x_2^R) \cdot E^{-2} 
(A^+ x_1^L + A^- x_2^R) \right) \,.
\eea

\subsection{Term linear in $a^{\dagger}$} \label{sub4}
From (\ref{start}), we see that the term linear in $a^{\dagger}$ is 
\bea
\ell &=& -2 i \, a^{\dagger} \cdot \left( \left( V_{12} \, D_{11} + V_{21} \, D_{21} \right) 
E^{-1} x_1 
+ \left( V_{12} \, D_{12} + V_{21} \, D_{22} \right) E^{-1} x_2
\right) \,,\ D \equiv \left({1 \over 1 + (1 - \eps)\Nu}\right)
\nonumber \\
&=& -2 i \, a^{\dagger} \cdot \left( \left( V_{12} \, D_{11} + V_{21} \, D_{21} \right) 
E^{-1} x_1 
+ C \left( V_{12} \, D_{11} + V_{21} \, D_{21} \right) C E^{-1} x_2 \right) \,.
\label{ax1}
\eea
We have
\bea
V_{12} \, D_{11} + V_{21} \, D_{21} &=& C Y \, D_{11} + Y C \, D_{21}
\nonumber \\
&=& {1 \over 1 - \eps} \, \left( C Y \, H^+(\eps) \, \left({1 \over 1 - \eps} + 
C X \right) \, H^-(\eps) - Y \, H^+(\eps) \, Z \, H^-(\eps) \right) \,.
\eea
Using (\ref{br1}) and (\ref{br2}), and noting that 
$\left( Y \, H^-_{\rm sing} \right)_{oo} 
= \left( Y \, H^-_{\rm sing} \right)_{oe} = 0$, 
we see that the singular terms don't contribute in $\ell$. it is then straightforward to 
show, using (\ref{EAAE}), that to zeroth order in $\eps$ we have 
\bea
&& V_{12} \, D_{11} + V_{21} \, D_{21} = \left( \begin{array}{cc} 
-\displaystyle{{1+U_{oo} \over 4-2 U_{oo}}} \;&\; \sqrt{3}i \, (4-2 U_{oo})^{-1} U_{oe} 
\crbig 
\half E^{-1} \Xgt_{eo} E \;&\; \half \end{array} \right) = 
\nonumber \\
&=& C E^{-1} \left( \begin{array}{cc} \half \;&\; \half (\Xgt_{eo})^{-1_L} \crbig 
\half \Xgt_{eo} \;&\; \half \end{array} \right) E + 
C E^{-1} \left( \begin{array}{cc} -E {1-2 U_{oo} \over 2-U_{oo}} E \;&\; 0 \crbig 0 \;&\; 0 
\end{array} \right) \left( \begin{array}{cc} \half \;&\; -\half (\Xgt_{eo})^{-1_L} \crbig 
-\half \Xgt_{eo} \;&\; \half \end{array} \right) E
\nonumber \\
&=& C E^{-1} A^+ \tilde{A}^+ E + C E^{-1} \Theta A^- \tilde{A}^- E \,, \ \ 
\Theta \equiv \left( \begin{array}{cc} -E {1-2 U_{oo} 
\over 2-U_{oo}} E \;&\; 0 \crbig 0 \;&\; 0 \end{array} \right) \,,
\label{ax2}
\eea
and thus 
\beq
C \left( V_{12} \, D_{11} + V_{21} \, D_{21} \right) C =  
C E^{-1} A^- \tilde{A}^- E + C E^{-1} \Theta A^+ \tilde{A}^+ E \,.
\label{ax3}
\eeq
Substituting (\ref{ax2}) and (\ref{ax3}) into (\ref{ax1}), we get
\bea
\ell &=& -2i \, a^{\dagger} \cdot C \left( E^{-1} A^+ x_1^L + E^{-1} 
\Theta A^- x_1^R + E^{-1} A^- x_2^R + E^{-1} \Theta A^+ x_2^L \right)
\nonumber \\
&=& -2i \, a^{\dagger} \cdot C E^{-1} \left(A^+ x_1^L + A^- X_2^R + \Theta A^- 
(x_1^R - x_2^L) \right)
\nonumber \\
&=& -2i \, a^{\dagger} \cdot C E^{-1} \left(A^+ x_1^L + A^- X_2^R \right) \,,
\eea
where in the second line we used $\Theta A^+ = \Theta C C A^+ = -\Theta A^-$, and in the 
third line we used the fact that the delta function forces $(x_1^R - x_2^L)$ to be zero.

\subsection{The overall normalization} \label{sub5}
It is now time to pause and look at what we have until now: 
\beq
\ket{x_1} \star \ket{x_2} = {\cal K}^{26}
\delta (x_1^R - x_2^L) \, 
\ket{A^+ x_1^L + A^- x_2^R} \,,
\eeq
where
\beq
{\cal K} \equiv \lim_{\eps \rightarrow 0} \, \left( 
K_0 \, \det (1 + (1 - \eps) \, \Nu)^{-\half} 
\, \det \left( {1 \over \pi \, \eps} \, M \right)^{-1/2} \right) \,,
\eeq
and $M$ is given by (\ref{M}).
We will now try to calculate ${\cal K}$ more explicitly. 
First note that the determinants appearing 
in ${\cal K}$ are not finite. To calculate them we use level truncation and 
keep oscillators of level $\leq N$ only, taking the 
limit $N \rightarrow \infty$ at the end. 
First, we have
\beq
\det \left( {1 \over \pi \, \eps} \, M \right)^{-1/2}  = 
\det (M)^{- \half} \, \eps^{N \over 2} \, \pi^{N \over 2} \,.
\eeq
Now note that $(1 + (1 - \eps) \, \Nu)^{-1}$ is a block matrix (of size $2 \, N$), 
but we can express its determinant in terms of the determinant of a $N$ by 
$N$ matrix:
\bea
(1 - \eps)^{2 N} \, \det \, (1 + (1 - \eps) \, \Nu)^{-1} &=& 
\det \, \left( \begin{array}{cc} 
H^+(\eps) \, ({1 \over 1 - \eps} + C X) \,  
H^-(\eps) \; & \; -H^+(\eps) \, Z \, H^-(\eps) \, C  
\crbig 
-C \, H^+(\eps) \, Z \, H^-(\eps) \; & \; 
C \, H^+(\eps) \, ({1 \over 1 - \eps} + C X) \, H^-(\eps) \, C \end{array} \right) 
\nonumber
\\
&=& \det \, \left( \begin{array}{cc} 
H^+(\eps) \; & \; -H^+(\eps) \, Z \, H^-(\eps) \, C  
\crbig 
C \, H^+(\eps) \; & \; 
C \, H^+(\eps) \, ({1 \over 1 - \eps} + C X) \, H^-(\eps) \, C \end{array} \right)
\nonumber
\\
&=& \det \, \left( \begin{array}{cc} 
H^+(\eps) \; & \; -H^+(\eps) \, Z \, H^-(\eps) \, C  
\crbig 
0 \; & \; 
C \, H^-(\eps) \, C \end{array} \right)
\nonumber
\\
&=& \det \left( H^+(\eps) \right)  \, \det \left( C \, H^-(\eps) \, C \right)
\nonumber
\\
& \stackrel{\eps \rightarrow 0}{\longrightarrow}&
\det \left( H^+(0) \right) \, \det \left( H^-_{\rm sing} \right) 
\, \eps^{-N} \,, 
\eea
where, in the second line we added, to the first $N$ columns, the last $N$ columns 
multiplied on the right by $C$; this operation doesn't change the determinant. Neither 
does the second operation: we subtracted the $N$ first lines, multiplied by $C$ on the 
left, to the $N$ last lines. This set to zero the lower-left block matrix, and the 
determinant then becomes the  product of the determinant of the upper-left block times 
the determinant of the lower-right block. In the last line we took the limit 
$\eps \rightarrow 0$ before the limit $N \rightarrow \infty$.
We can now write 
\beq
{\cal K} = K_0 \, \det \left( H^+(0) \right)^\half \, \det (E) \, \det(A^-)^{-1} \, 
\pi^{N \over 2} \,.
\eeq
We see that the $\eps$'s canceled each other, proving that our regularization scheme 
is consistent.

\paragraph{}
This normalization constant ${\cal K}$ plays a role when we want to express the 
star-product in the space of wavefunctions $\psi(x^L, x^R)$. We define the 
wavefunction of the state $\ket{\psi}$ by: $\psi(x^L, x^R) \equiv \bra{x} \psi \rangle$. 
We expect the star-product of wavefunctions to have the following form:
\beq
(\psi_1 \star \psi_2)(x^L, x^R) = {\cal Z} \, \int [d y] \, \psi_1(x^L, y) \, 
\psi_2(y, x^R) \,,
\label{wavestar}
\eeq
where we integrate over the half-string $y$. In \cite{RSZhalf}, ${\cal Z}$ was absorbed 
in the integration measure\footnote{I thank B.~Zwiebach for a discussion about this 
point} $[d y]$, but here we want to define the measure in the obvious way: 
$[d y] = \prod{d y_n}$, as a product over all half-string mode measures. 
And similarly for the path integral over whole-string modes: $[d x] = \prod{d x_n}$.

\paragraph{}
In order to formally prove (\ref{wavestar}) we calculate 
$\ket{\psi_1} \star \ket{\psi_2}$ by inserting two complete sets of zero-momentum states. 
And we will need to write $[d x] = J \, [d x^L] [d x^R]$, where $J$ is some constant 
relating the measures $[d x]$ and $[d x^L] [d x^R]$. 
\bea
\ket{\psi_1} \star \ket{\psi_2} &=& \left(\int [d x_1] \ket{x_1} \bra{x_1} \psi_1 \rangle 
\right) \star \left(\int [d x_2] \ket{x_2} \bra{x_2} \psi_1 \rangle \right)
\nonumber
\\
&=& \int [d x_1] [d x_2] \, \psi_1(x_1^L, x_1^R) \,  \psi_2(x_2^L, x_2^R) \,
\left( \ket{x_1} \star \ket{x_2} \right) 
\nonumber
\\
&=& J^2 {\cal K}^{26} \int [d x_1^L] [d x_1^R] [d x_2^L] [d x_2^R] \, 
\psi_1(x_1^L, x_1^R) \,  \psi_2(x_2^L, x_2^R) \, \delta(x_1^R - x_2^L) \, 
\ket{A^+ x_1^L + A^- x_2^R}
\nonumber
\\
&=& J^2 {\cal K}^{26} \int [d x^L] [d x^R] \left( \int [d y] \, \psi_1(x^L, y) \, 
\psi_2(y, x^R) \right) \ket{x}
\nonumber
\\
&=& J {\cal K}^{26} \int [d x] \left( \int [d y] \, \psi_1(x^L, y) \, 
\psi_2(y, x^R) \right) \ket{x} \,.
\eea
Therefore, if we impose the following normalization on the states $\ket{x}$:
\beq
\int [d x] \ket{x} \bra{x} = 1 \,,
\eeq
we get
\beq
{\cal Z} = J {\cal K}^{26} \,.
\eeq

\paragraph{}
We postpone to future work the exact calculation of the factors\footnote{
It must be possible to relate these factors to the constants $\gamma_1$, 
$\gamma_2$ and $\gamma_3$ calculated in Appendices A.1 and A.2 of 
\cite{GT2}. I thank W.~Taylor for pointing this out.} ${\cal Z}$ or $J$.

\section{Summary} \label{s4}

In this summary, we would like to collect some useful formulas that were 
established in this paper.

\subsection{Factorization of the sliver}

\beq
A^{+T} E^{-1} {1 - S \over 1 + S} E^{-1} A^- = 0 \,,
\eeq
\bea
\sqrt{3} \, i \, U_{eo} \, (1+U_{oo})^{-1} = E^{-1} \, \Xgt_{eo} \, E \,,
\crbig
-\sqrt{3} \, i \, (1 - U_{oo}) \, (U_{eo})^{-1} = E \, \Xgt_{oe} \, E^{-1} \,.
\eea

\subsection{Star-product}

If $\ket{x_1}$ and $\ket{x_2}$ are zero-momentum eigenstates of $\hat{x}$, then: 
\beq
\ket{x_1} \star \ket{x_2} = {\cal K}^{26}
\delta (x_1^R - x_2^L) \, 
\ket{A^+ x_1^L + A^- x_2^R} \,,
\eeq
where
\beq
{\cal K} = K_0 \, \det \left( H^+(0) \right)^\half \, \det (E) \, \det(A^-)^{-1} \, 
\pi^{N \over 2} \,.
\eeq

The star-product of two zero-momentum string wavefunctions $\psi_1$ and $\psi_2$ is: 
\beq
(\psi_1 \star \psi_2)(x^L, x^R) = J {\cal K}^{26} \, \int [d y] \, \psi_1(x^L, y) \, 
\psi_2(y, x^R) \,,
\eeq
where $J$ is defined by
\beq
[d x] = J \, [d x^L] [d x^R] \,.
\eeq

\subsection{Projectors on the sliver}

\beq
\rho_1 \, (1+S) \, E \, \tilde{A}^{+T} = 0 \,, \qquad 
\rho_2 \, (1+S) \, E \, \tilde{A}^{-T} = 0 \,.
\eeq

\section{Discussion and Conclusions} \label{s5}

With an algebraic proof, very similar to the one in \cite{GT}, we were able to show that 
the sliver wavefunction factorizes into a left part and a right part. This was 
already strongly believed to be true on both geometrical and numerical grounds, 
and it was moreover known to be true for the D-instanton sliver. This property 
is very important in generating classical solutions in Vacuum String Field Theory. 
And we already found an application of this factorization, that we used in proving 
that the matrices $\rho_1$ and $\rho_2$ are left and right projectors on the sliver.

\paragraph{}
We have performed a complete analytic calculation of the star-product of two 
eigenstates of $\hat{x}$ expressed in the oscillator basis, and have shown that 
we do get the expected result. That is: The star-product is zero unless the 
right-half of the first string exactly coincides with the left-half of the second 
string. Moreover, the product is an eigenstate of $\hat{x}$, whose left-half is the 
left-half of the first string, and whose right-half is the right-half of the 
second string. This allowed us to write the star-product between zero-momentum 
string wavefunctions as a path integral over a half-string, modulo a still unknown 
constant (which can be absorbed in the half-string integration measure).

\paragraph{}
This calculation, done with the infinite dimensional matrix representation of the 
Neumann coefficients was technically difficult, partly because we had to 
find inverses to nontrivial block-matrices and because we had to use a regulator 
in order to make singular matrices invertible. We hope that the techniques used here 
will find some other useful applications in performing exact calculations in 
string field theory.

\paragraph{}
A direct extension of this work would be to generalize the calculation of the 
star-product of zero-momentum eigenstates of $\hat{x}$ to include nonzero-momentum 
eigenstates as well. Also, it would be interesting to generalize these results to the 
ghost sector. At last, it might be 
useful to be able to relate the integration measure over the whole-string modes to 
the measure over the half-string modes.


\section*{Acknowledgements}
I would like to thank B.~Zwiebach for suggesting to me some of the problems treated 
in this paper, as well as for many helpful discussions, and proofreadings of the 
manuscript. I also thank 
I.~Ellwood, B.~Feng, Y.-H.~He and W.~Taylor for discussions.
This research was supported in part by
the CTP and LNS of MIT, a Presidential Fellowship of MIT 
and the U.S. Department of Energy 
under cooperative research agreement \# DE-FC02-94ER40818.


\bibliographystyle{plain}

\end{document}